\documentclass[prb,preprint,showpacs,preprintnumbers,amsmath,amssymb]{revtex4}

\usepackage{graphicx,subfigure}
\usepackage{dcolumn}
\usepackage{bm}
\usepackage{psfrag}
\usepackage{multirow}

\def\be{\begin{equation}}       \def\ee{\end{equation}}
\def\bea{\begin{eqnarray}}      \def\eea{\end{eqnarray}}
\def\half{\frac{1}{2}}
\def\dag{\dagger}
\def\non{\nonumber}
\begin{document}

\title{Magnetic phase diagram of spatially anisotropic, frustrated spin-$\half$ Heisenberg antiferromagnet on a stacked square lattice}
\author{Kingshuk Majumdar}
\affiliation{Department of Physics, Grand Valley State University, Allendale, 
Michigan 49401, USA}
\email{majumdak@gvsu.edu}
\date{\today}

\begin{abstract}\label{abstract}
Magnetic phase diagram of a spatially anisotropic, frustrated spin-$\half$ Heisenberg antiferromagnet on a stacked square lattice is investigated using second-order spin-wave expansion. The effects of interlayer
coupling and the spatial anisotropy on the magnetic ordering of two ordered ground states are explicitly studied. It is shown that with increase in next nearest neighbor frustration the second-order corrections play a significant role in stabilizing the magnetization. We obtain two ordered magnetic phases (Ne\'{e}l and stripe) separated by a paramagnetic disordered phase.  Within second-order spin-wave expansion we find  that the width of the disordered phase diminishes with increase in the interlayer coupling or with decrease in spatial anisotropy but it does not disappear. Our obtained phase diagram differs significantly from the phase diagram obtained using linear spin-wave theory.
\end{abstract}

\maketitle

\section{\label{sec:Intro}Introduction}
Availability of new magnetic materials and the recent discovery of superconductivity at relatively high
temperatures in the iron pnictide family of materials have spurred a flurry of interest in understanding
the properties of frustrated magnets.~\cite{kim99,coldea01,ronnow01,chris04,chris07,bombardi04,melzi00,melzi01,carretta02,rosner02, kamihara08,cruz08,klaus08,Dong08} For the last two decades the properties of quantum spin-$\half$ Heisenberg antiferromagnet (HAFM) with nearest neighbor (NN) $J_1$ and next nearest neighbor exchange interactions (NNN) $J_2$ on a square lattice have been studied extensively by various analytical and numerical 
techniques.~\cite{harris71,chandra88,sudip89,castilla91,canali92A,igar92,igar93,majumdar10b,
caprioti03,ein91,ein92,dot94,valeri99,nersesyan03,oleg04,shannon04,isaev09,weihong91,hamer92,oitmaa96,weihong05,sindzingre04,white07,tsirlin09,sandvik01,capriotti03,yunoki04} It is now well established that at low temperatures these systems exhibit new types of magnetic order and novel quantum phases.~\cite{diep,subir1} For $J_2=0$ the ground state is antiferromagnetically ordered at low temperatures. Addition of NNN interactions induces
a strong frustration and break the antiferromagnetic (AF) order. The competition between 
NN and NNN interactions for the square lattice is characterized by the frustration parameter 
$\eta= J_2/J_1$. A disordered paramagnetic phase probably columnar dimer exists between 
$\eta_{1c} \approx 0.38$ and $\eta_{2c} \approx 0.60$. 
For $\eta<\eta_{1c}$ the square lattice is AF-ordered whereas for $\eta>\eta_{2c}$ a 
degenerate collinear antiferromagnetic (CAF) stripe phase emerges. Experimentally by applying high pressure the ground state phase diagram of these frustrated spin systems can be explored from low $\eta=J_2/J_1$ to high $\eta$. For example in Li$_2$VOSiO$_4$ X-ray diffraction measurements show that the value of $\eta$ decreases by about 40\% with increase in pressure from zero to 7.6 GPa.~\cite{pavarini08} Moreover, nuclear magnetic resonance, magnetization, specific heat, and 
muon spin rotation measurements on the compounds Li$_2$VOSiO$_4$, Li$_2$VOGeO$_4$, VOMoO$_4$, and BaCdVO(PO$_4$)$_2$ show significant coupling between NN and NNN neighbors.~\cite{melzi00,melzi01,bombardi04} In addition these experiments on Li$_2$VOSiO$_4$ have shown that it undergoes a phase transition at a low temperature (2.8 K) to collinear AF order with magnetic moments lying in the $a-b$ plane  with $J_2+J_1 \sim 8.2(1)$ K 
and $J_2/J_1 \sim 1.1 (1)$.~\cite{melzi01,carretta02}

A generalization of the frustrated $J_1-J_2$ model is the $J_1-J_1^\prime-J_2$ model where 
$\zeta=J_1^\prime/J_1$ is the directional anisotropy parameter.~\cite{nersesyan03,oleg04,majumdar10b}  A possible candidate of this model may be the compound (NO)Cu(NO$_3$)$_3$.~\cite{volkova10} Extensive
band structure calculations~\cite{tsirlin09} for the vanadium phosphate compounds Pb$_2$VO(PO$_4$)$_2$, SrZnVO(PO$_4$)$_2$, BaZnVO(PO$_4$)$_2$, BaCdVO(PO$_4$)$_2$ have shown four different exchange couplings: $J_1$ and $J_1^\prime$ between the NN and $J_2$ and $J_2^\prime$ between NNN. For example $\zeta \approx 0.7$ and $J_2^\prime/J_2 \approx 0.4$ were obtained for SrZnVO(PO$_4$)$_2$. Recently using second-order spin-wave expansion the effects of directional anisotropy on the spin-wave energy dispersion, renormalized spin-wave velocities, and magnetizations for the two ordered phases have been studied in detail.~\cite{majumdar10b} It has been found that the spatial anisotropy reduces the width of the disordered phase.

Although much efforts have been made to understand the properties of two dimensional (2D) frustrated magnets
research on three dimensional (3D) systems has been limited.~\cite{oitmaa,lallemand,azaria,derrida1,derrida2,oguchi,katanin,banavar,majumdar09,
majumdar10a} Earlier work on HAFM on the pyrochlore lattice~\cite{canals98} and on the stacked kagome lattice~\cite{Fak08SL,subir92kagome,harris92Kagome,chubu92Kagome} showed existence of a magnetically disordered phase. On the contrary, spin-wave calculations for the 3D $J_1-J_2$ model on the body-centered cubic (BCC) lattice and for the simple cubic lattice (SC) show no signs of an intermediate disordered paramagnetic phase.\cite{schmidt02,majumdar09,majumdar10a} Instead for the BCC and the SC lattice a direct first order phase transition occurs from a two-sublattice N\'{e}el ordered AF phase for 
small $J_2$ to a collinear antiferromagnetic ordered state for large $J_2$.\cite{majumdar09,majumdar10a} 
For Li$_2$VOSiO$_4$, a layered material that can be described by a square lattice $J_1-J_2$ model with large $J_2$ the interlayer  coupling $J_\perp/J_1 \sim 0.07$ is not negligible.\cite{rosner02} Due to a finite interlayer magnetic coupling $J_\perp$ these experimental systems are quasi 2D.

Another example is the recently discovered iron pnictide superconductors.\cite{kamihara08} The parent phases of these materials have been found to be
metallic but with AF order and magnetic excitations have shown to play an important role in the superconducting state.\cite{kamihara08,cruz08,klaus08,Dong08,apple10,singh09,uhrig09,yao08,yao10} Although magnetism in these materials
are still debated neutron scattering spectra for the pnictides show sharp spin-waves.\cite{zhao08} 
These studies have also revealed that the parent compounds exhibit a columnar antiferromagnetic ordering
with a staggered magnetic moment of $(0.3-0.4)\mu_B$ in LaOFeAs and $(0.8-1.01)\mu_B$ in 
Sr(Ba,Ca)Fe$_2$As$_2$.\cite{Dong08,zhao08} Moreover, at low temperatures 
there is orthorhombic distortion and the exchange constants have been found to be substantially 
anisotropic.\cite{diallo09,zhao09}
Motivated by the observation of spatially anisotropic exchange constants in these materials the spin-wave
spectra and the low-temperature phases of the model can be studied by the spatially anisotropic 
$J_1-J_1^\prime-J_2$ Heisenberg model on a square lattice with NN exchanges $J_1$ along the $x$ axis, $J_1^\prime$ along the $y$ axis, and NNN interactions $J_2$ along the diagonals in the $xy$ plane. Recent experiments on iron-based superconductors such as undoped iron oxypnictides reveal that the electronic couplings are more three dimensional than in the cuprate superconductors.\cite{zhao08,ewings08,yuan09} With decrease in
temperature most undoped iron-pnictide superconductors show a structural transition from a tetragonal paramagnetic phase to a orthorhombic phase. In the 122 materials a three dimensional long-range
antiferromagnetic order develops simultaneously. 

These quasi 2D frustrated layered systems with anisotropic magnetic exchange couplings and with non-negligible interlayer couplings $J_\perp$ serve as a motivation to investigate the $J_1-J_1^\prime-J_2-J_\perp$ model. This model on a square lattice with isotropic NN interactions ($J_1=J_1^\prime$) has been studied numerically 
by coupled-cluster and the rotation-invariant Green's function methods.\cite{darradi06} These calculations show that the quantum paramagnetic 
phase disappears for a critical value of interlayer coupling $J^c_\perp \approx 0.23 J_1$. For $J_\perp <J^c_\perp$ a
second-order phase transition between the quantum Ne\'{e}l to a quantum paramagnetic phase and then a first-order  transition from the quantum paramagnetic phase to the stripe phase occur. For $J_\perp>J_\perp^c$ there is a direct first-order transition between the Ne\'{e}l to the stripe phase. Existence of this critical point has also
been reported by other authors where they have used effective field theory in a finite cluster
to obtain $J^c_\perp \approx 0.67J_1$.\cite{nunes10} In the context of iron pnictides this model has been studied using
linear spin-wave theory (LSWT) to obtain the spin-wave energy dispersion and the sublattice magnetization.~\cite{yao10} However, to our knowledge this model has not been studied using higher order
spin-wave expansion. We will find that second-order corrections due to quantum fluctuations increase substantially as the classical phase transition point is approached. As a result the magnetic phase
diagram obtained from second-order spin-wave expansion differs significantly from the phase diagram
obtained by LSWT.

In this work we investigate the magnetic phase diagram of the 
$J_1-J_1^\prime-J_2-J_\perp$ Heisenberg AF on a stacked square lattice. We use spin-wave expansion based on Holstein-Primakoff transformation up to second-order in $1/S$ to numerically calculate the sublattice magnetization for each of the two ordered magnetic phases. 
The paper is organized as follows. Section~\ref{sec:model} provides an introduction
to the Hamiltonian for the Heisenberg spin-$\half$ AF on a spatially 
anisotropic stacked square lattice. The classical ground state
configurations of the model and the two phases are then briefly discussed. In the next
two sections Sec.~\ref{sec:AFphase} and Sec.~\ref{sec:CAFphase} the spin Hamiltonian 
is mapped to the Hamiltonian of interacting spin-wave excitations and 
spin-wave expansion for sublattice magnetizations are presented for the two phases. Magnetizations for the two phases are numerically 
calculated with different values of interlayer coupling and spatial anisotropy and the results are plotted and discussed in Section~\ref{sec:results}. 
Finally we summarize our results in Section~\ref{sec:conclusions}.

\section{\label{sec:model} Theory}
We consider a spatially anisotropic, frustrated spin-$\half$ HAFM on a  $N_L \times N_L \times 
N_L$ cubic lattice with four types
of exchange interactions between spins: $J_1$ along the $x$ (row) 
direction, 
$J_1^\prime$ along the $y$ (column) direction, $J_2$ along the diagonals in the $xy$ plane, and 
$J_\perp$ is the interlayer coupling. We assume all interactions to be AF 
and positive i.e. $J_1,J_1^\prime,J_2,J_\perp >0$. This $J_1-J_1^\prime-J_2-J_\perp$ spin system is 
described by the Heisenberg Hamiltonian
\be
H = \half \sum_{i, \ell} \Big[J_1{\bf S}_{i,\ell} \cdot {\bf S}_{i+\delta_x,\ell}
+ J_1^\prime {\bf S}_{i,\ell} \cdot {\bf S}_{i+\delta_y,\ell} 
+ J_2 {\bf S}_{i,\ell} \cdot {\bf S}_{i+\delta_x + \delta_y,\ell}\Big]
+\half J_\perp \sum_{i,\ell}{\bf S}_{i,\ell} \cdot {\bf S}_{i,\ell+1},
\label{hamiltonian}
\ee
where $\ell$ labels the layers, $i$ runs over all lattice sites and $i+\delta_x$ ($\delta_x =\pm 1$) 
and $i+\delta_y$ ($\delta_y =\pm 1$) are the 
nearest neighbors to the $i$-th site along the row and the column direction. The third
term represents the interaction between the next-nearest neighbors, which are along the
diagonals in the $xy$ plane and the last term is for the NN coupling between the layers. This model is different from the fully frustrated simple cubic lattice which has the additional 
NNN interactions in the $xz$ and $yz$ planes.~\cite{majumdar10a}

At zero temperature this model exhibits three types of classical ground state (GS)
configurations: the Ne\'{e}l  or the ($\pi,\pi, \pi$) state and the two stripe states 
which are the 
columnar stripe ($\pi,0,\pi$) and the row stripe ($0,\pi,\pi$). The classical ground state energies of these states are 
\bea
E^{\rm class}_{\rm AF}/N &=& -\half J_1S^2 z \left[1+\zeta - 2 \eta -\delta\right], \nonumber \\
E^{\rm class}_{\rm col}/N &=& -\half J_1S^2 z \left[1-\zeta + 2 \eta+\delta\right],\label{cgs}\\
E^{\rm class}_{\rm row}/N &=& -\half J_1S^2 z \left[-1+\zeta + 2 \eta+\delta\right],\nonumber
\eea
where $\zeta = J_1^\prime/J_1$ measures the directional
anisotropy, $\eta=J_2/J_1$ is the magnetic frustration between the 
NN (row direction) and NNN spins, and $\delta=J_\perp/J_1$ is the interlayer coupling 
parameter. $z=2$ is the number of NN sites along the row (column) direction. For $\eta<\zeta/2$ 
the classical GS is the AF
Ne\'{e}l state and for $\eta>1/2$ (with $\zeta=1$) the GS is doubly degenerate. Otherwise for $\zeta<1$ the GS is the columnar AF ($\pi,0,\pi$) state.
The classical first-order phase transition between the AF and CAF state
occurs at the critical value $\eta_c^{\rm class}= \zeta/2$, which is independent of $\delta$. 

The motivation of this paper is to investigate the role of interlayer coupling $\delta$ and spatial 
anisotropy $\zeta$ to the quantum phases 
of this model. How does the quantum fluctuations due to $\delta$ and $\zeta$ affect the disordered paramagnetic phase? Is there a critical point for $\delta$ (or $\zeta$) above (or below) which the intermediate quantum paramagnetic GS does not exist and we have a direct first-order phase transition from  the AF to the CAF ordered phase? For our study we follow the standard procedure by
first expressing the fluctuations around the ``classical" ground state in terms of the
boson operators using the Holstein-Primakoff transformation. The quadratic term in the boson
operators corresponds to the LSWT, whereas the higher-order terms 
represent spin-wave (magnon) interactions. We keep terms up to second order in $1/S$. 
The staggered magnetization per spin to the leading order in $1/S^2$ for the AF and CAF phases are then obtained from the renormalized magnon Green's functions and self-energies. We will follow the theoretical framework described in detail in Ref.~\onlinecite{majumdar10b}. However for completeness we provide the necessary equations that are required for numerical computations and to follow the present work.
\subsection{\label{sec:AFphase} ($\pi,\pi,\pi$) AF Ne\'{e}l Phase}
For the AF ordered phase NN interactions are between A and B sublattices and 
NNN interactions are between A-A and B-B sublattices. The Hamiltonian in Eq.~\ref{hamiltonian}
takes the form:
\bea
H &=& J_1 \sum_{i,\ell}{\bf S}_{i,\ell}^{\rm A} \cdot {\bf S}_{i+\delta_x,\ell}^{\rm B}
+ J_1^\prime \sum_{i,\ell}{\bf S}_{i,\ell}^{\rm A} \cdot {\bf S}_{i+\delta_y,\ell}^{\rm B} 
+ \half J_2 \sum_{i,\ell}\Big[ {\bf S}_{i,\ell}^{\rm A} \cdot {\bf S}_{i+\delta_x + \delta_y,\ell}^{\rm A}
+ {\bf S}_{i,\ell}^{\rm B} \cdot {\bf S}_{i+\delta_x + \delta_y,\ell}^{\rm B}\Big]\nonumber \\
&+& J_\perp \sum_{i,\ell}{\bf S}_{i,\ell}^{\rm A} \cdot {\bf S}_{i,\ell+1}^{\rm B}.
\label{ham-AF}
\eea
This Hamiltonian is mapped into an equivalent Hamiltonian
of interacting bosons by transforming the spin operators to bosonic creation 
and annihilation operators $a^\dag, a$ for ``up''  and $b^\dag, b$ for ``down'' sublattices using the 
Holstein-Primakoff transformations keeping only terms up to the order of $1/S^2$
\begin{eqnarray}
S_{i,\ell}^{A+} &\approx& \sqrt{2S}\Big[1- \half \frac {a_{i\ell}^\dag a_{i\ell}}{(2S)}
-\frac 1{8} \frac{a_{i\ell}^\dag a_{i\ell} a_{i\ell}^\dag a_{i\ell}}{(2S)^2} \Big]a_{i\ell},\non \\
S_{i,\ell}^{A-} &\approx& \sqrt{2S}a_{i\ell}^\dag \Big[1-\half \frac {a_{i\ell}^\dag a_{i\ell}}{(2S)} 
-\frac 1{8} \frac{a_{i\ell}^\dag a_{i\ell} a_{i\ell}^\dag a_{i\ell}}{(2S)^2} \Big], \non \\
S_{i,\ell}^{Az} &=& S-a^\dag_{i\ell}a_{i\ell},  \label{holstein}  \\ 
S_{j,\ell}^{B+} &\approx& \sqrt{2S}b_{j\ell}^\dag \Big[1-\half \frac {b_{j\ell}^\dag b_{j\ell}}{(2S)} 
-\frac 1{8} \frac{b_{j\ell}^\dag b_{j\ell} b_{j\ell}^\dag b_{j\ell}}{(2S)^2}\Big],\non \\
S_{j,\ell}^{B-} &\approx& \sqrt{2S}\Big[1-\half \frac {b_{j\ell}^\dag b_{j\ell}}{(2S)}
 -\frac 1{8} \frac{b_{j\ell}^\dag b_{j\ell} b_{j\ell}^\dag b_{j\ell}}{(2S)^2} \Big]b_{j\ell}, \non \\
S_{j,\ell}^{Bz} &=& -S+b^\dag_{j\ell}b_{j\ell}. \non 
\end{eqnarray}
In powers of $1/S$ the Hamiltonian is now written as
\be
H = -\half N J_1 S^2 z(1+\zeta)\Big[1 - \frac {2\eta}{1+\zeta}\Big] + H_0 + H_1+H_2 + \ldots.
\ee 
The first term corresponds to the classical energy of the AF ground state (Eq.~\ref{cgs}).
Then the real space Hamiltonian is transformed to the ${\bf k}$-space Hamiltonian. Momentum
${\bf k}$ is defined in the first Brillouin zone (BZ): $-\pi < k_x \leq \pi,\; 
-\pi <k_y \leq \pi ,\; -\pi < k_z \leq \pi \;$ (with unit lattice spacings). 
Next we diagonalize the quadratic part $H_0$ by transforming the 
operators $a_{\bf k}$ and $b_{\bf k}$ to magnon operators 
$\alpha_{\bf k}$ and $\beta_{\bf k}$ using the  Bogoliubov (BG) transformations
\be
a^\dag_{\bf k} =l_{\bf k} \alpha_{\bf k}^\dag + m_{\bf k}\beta_{-{\bf k}},\;\;\;
b_{-\bf k} =m_{\bf k} \alpha_{\bf k}^\dag + l_{\bf k}\beta_{-{\bf k}},
\ee
where the coefficients $l_{\bf k}$ and $m_{\bf k}$ are defined as
\be
l_{\bf k} = \Big[\frac {1+\epsilon_{\bf k}}{2\epsilon_{\bf k}} \Big]^{1/2},\;\;
m_{\bf k} = -{\rm sgn}(\gamma_{\bf k})\Big[\frac {1-\epsilon_{\bf k}}
{2\epsilon_{\bf k}} \Big]^{1/2}\equiv -x_{\bf k}l_{\bf k},
\ee
with 
\bea
\epsilon_{\bf k} &=& (1-\gamma_{\bf k}^2)^{1/2}, \non \\
\gamma_{\bf k} &=& \gamma_{1{\bf k}}/\kappa_{\bf k}, \non \\
\gamma_{1{\bf k}}&=&[\cos (k_x)+\zeta \cos (k_y)+\delta \cos(k_z)]/(1+\zeta), \label{defs-AF} \\
\gamma_{2{\bf k}} &=& \cos (k_x)\cos(k_y), \non \\
\kappa_{\bf k} &=& 1- \frac {2\eta}{1+\zeta} (1-\gamma_{2{\bf k}})+\frac{\delta}{1+\zeta}.\non
\eea
$\gamma_{\bf k}$ is 
negative in certain parts of the first BZ - so it is essential to keep track of the sign
of $\gamma_{\bf k}$ through the function ${\rm sgn} (\gamma_{\bf k})$.
After these transformations, the quadratic part of the Hamiltonian becomes
\be
H_0 = J_1Sz(1+\zeta)\sum_{\bf k} \kappa_{\bf k}\left(\epsilon_{\bf k} -1\right)
+J_1Sz(1+\zeta)\sum_{\bf k}\kappa_{\bf k} \epsilon_{\bf k}
\left( \alpha^\dag_{\bf k}\alpha_{\bf k}+\beta^\dag_{\bf k}\beta_{\bf k}\right).
\label{H0term}
\ee
The first term is the zero-point energy and the second term represents the 
excitation energy of the magnons within LSWT.

The part $H_1$ corresponds to $1/S$ correction to the Hamiltonian. We follow the same
procedure as described above. The resulting expression after
transforming the bosonic operators to the magnon operators is
\bea
H_1 &=& \frac {J_1Sz(1+\zeta)}{2S}\sum_{\bf k}
\Big[ A_{\bf k}\left(\alpha^\dag_{\bf k}\alpha_{\bf k}+
\beta^\dag_{\bf k}\beta_{\bf k}\right) 
+ B_{\bf k}\left(\alpha^\dag_{\bf k}\beta_{-\bf k}^\dag+
\beta_{-\bf k}\alpha_{\bf k}\right)\Big] \non \\
&-& \frac {J_1Sz(1+\zeta)}{2SN}\sum_{1234}
\delta_{\bf G}(1+2-3-4)l_1l_2l_3l_4\Big[\alpha_1^\dag \alpha_2^\dag \alpha_3 \alpha_4
V_{1234}^{(1)} +\beta^\dag_{-3}\beta^\dag_{-4}\beta_{-1}\beta_{-2}V_{1234}^{(2)} \non \\
&+&4\alpha_1^\dag \beta_{-4}^\dag \beta_{-2}\alpha_3 V_{1234}^{(3)} +\Big\{
2\alpha_1^\dag \beta_{-2}\alpha_3 \alpha_4V_{1234}^{(4)} +2\beta_{-4}^\dag \beta_{-1}
\beta_{-2}\alpha_3 V_{1234}^{(5)} + \alpha_1^\dag \alpha_2^\dag \beta_{-3}^\dag 
\beta_{-4}^\dag V_{1234}^{(6)} \non \\
&+& h.c.\Big\}\Big].
\label{H1term}
\eea 
In the above equation three-dimensional momenta ${\bf k}_1, {\bf k}_2, {\bf k}_3, {\bf k}_4$ are abbreviated
as 1, 2, 3, and 4. The first term in Eq.~\ref{H1term} is obtained by setting the products
of four boson operators into normal ordered forms with respect to the magnon
operators, where $A_{\bf k}$ and $B_{\bf k}$ are 
\bea
A_{\bf k}&=& A_1 \frac 1{\kappa_{\bf k}\epsilon_{\bf k}}\Big[\kappa_{\bf k}
-\gamma_{1{\bf k}}^2\Big] + A_2 \frac 1{\epsilon_{\bf k}}
\Big[1-\gamma_{2{\bf k}}\Big], \\
B_{\bf k} &=& B_1 \frac {1}{\kappa_{\bf k}\epsilon_{\bf k}}
\gamma_{1{\bf k}}\Big[1-\gamma_{2{\bf k}}\Big],
\eea
with
\bea
A_1 &=& \frac 2{N} \sum_{\bf p} \frac 1{\epsilon_{\bf p}}
\Big[\frac {\gamma_{1{\bf p}}^2}{\kappa_{\bf p}}+\epsilon_{\bf p}-1\Big], \\
A_2 &=& \Big(\frac {2\eta}{1+\zeta} \Big)\frac 2{N} \sum_{\bf p} 
\frac 1{\epsilon_{\bf p}}\Big[1-\epsilon_{\bf p}-\gamma_{2{\bf p}}\Big], \\
B_1 &=& \Big(\frac {2\eta}{1+\zeta} \Big)\frac 2{N} \sum_{\bf p} 
\frac 1{\epsilon_{\bf p}}\Big[\gamma_{2{\bf p}}-
\frac {\gamma_{1{\bf p}}^2}{\kappa_{\bf p}}\Big].
\eea
The second term in Eq.~\ref{H1term} represents scattering between spin-waves where the three-dimensional delta
function $\delta_{\bf G}(1+2-3-4)$ ensures that momentum is conserved within a 
reciprocal lattice vector ${\bf G}$. Explicit
forms of the vertex factors $V_{1234}^{i=1...6}$ are given in Ref.~\onlinecite{majumdar10b}.
Here we provide two of the vertex factors that are needed to calculate the magnetization. They are
\bea
V_{1234}^{(4)} &=&-\gamma_1 (2-4)x_4-\gamma_1(1-4)x_1x_2x_4-\gamma_1(2-3)x_3
- \gamma_1(1-3)x_1x_2x_3  \non \\
&+& \half \Big[\gamma_1(2) +\gamma_1 (1) x_1x_2 +\gamma_1(3) x_2x_3+
\gamma_1(4)x_2x_4 \non \\
&+&\gamma_1 (2-3-4)x_3x_4
+ \gamma_1(1-3-4)x_1x_2 x_3x_4+ \gamma_1(3-2-1)x_1x_3+\gamma_1(4-2-1)x_1x_4\Big]\non \\
&+& \Big(\frac {2\eta}{1+\zeta}\Big) f_{1234}\Big[x_2+{\rm sgn}(\gamma_{\bf G})x_1x_3x_4
\Big],\\
V_{1234}^{(6)} &=&\gamma_1 (2-4)x_2x_3+\gamma_1(2-3)x_2x_4+\gamma_1(1-3)x_1x_4
+ \gamma_1(1-4)x_1x_3  \non \\
&-& \half \Big[\gamma_1(2)x_2x_3x_4 +\gamma_1 (3)x_4 +\gamma_1(2-3-4)x_2+
\gamma_1(3-2-1)x_1x_2x_4 \non \\
&+&\gamma_1 (1)x_1x_3x_4
+ \gamma_1(4) x_3+ \gamma_1(1-3-4)x_1+\gamma_1(4-2-1)x_1x_2x_3\Big]\non \\
&-& \Big(\frac {2\eta}{1+\zeta}\Big) f_{1234}\Big[x_3x_4+{\rm sgn}(\gamma_{\bf G})x_1x_2
\Big],
\eea
with
\be
f_{1234}=\half\Big[\gamma_2(1-3)+\gamma_2(1-4)+\gamma_2(2-3)+\gamma_2(2-4)
-\gamma_2(1)-\gamma_2(2)-\gamma_2(3)-\gamma_2(4) \Big].
\ee

The second order term, $H_2$ is composed of six boson operators. After transformation to magnon operators 
$\alpha_{\bf k},\beta_{\bf k}$ the Hamiltonian
in normal ordered form reduces to
\be
H_2= \frac {J_1Sz(1+\zeta)}{(2S)^2} \sum_{\bf k} 
\Big[ C_{1{\bf k}}\left(\alpha^\dag_{\bf k}\alpha_{\bf k}+\beta^\dag_{\bf k}\beta_{\bf k}
\right)+C_{2{\bf k}}\left(\alpha^\dag_{\bf k}\beta_{-\bf k}^\dag+
\beta_{-\bf k}\alpha_{\bf k}\right)+\ldots \Big].
\label{H2term}
\ee
The dotted terms contribute to higher than second-order corrections and are thus omitted in 
our calculations.  
The coefficients $C_{1{\bf k}}$ and $C_{2{\bf k}}$ are 
\bea
C_{1{\bf k}} &=& \half l_k^2 \Big(\frac {2}{N}\Big)^2 \sum_{12} l_1^2l_2^2\Big[
-6\gamma_1(2-1-{\bf k})x_{\bf k}x_1x_2 + \gamma_1(2) x_1^2x_2+
\gamma_1(2)x_{\bf k}^2x_1^2x_2 \non \\
&+& 2\gamma_1({\bf k})x_{\bf k}x_1^2 
+\gamma_1(1)x_{\bf k}^2x_1+\gamma_1(2)x_2\Big]
- \frac 1{4}\Big(\frac {2\eta}{1+\zeta} \Big) l_{\bf k}^2(1+x_{\bf k}^2)
{\tilde C}_{\bf k}, \\
C_{2{\bf k}} &=& \half l_k^2 \Big(\frac {2}{N}\Big)^2 \sum_{12} l_1^2l_2^2\Big[
3\gamma_1(2-1-{\bf k})x_1x_2 + 3\gamma_1(2-1-{\bf k}) x_{\bf k}^2x_1x_2-
2\gamma_1(1)x_{\bf k}x_1x_2^2 \non \\
&-& 2\gamma_1(2)x_{\bf k}x_2 
-\gamma_1({\bf k})x_2^2-\gamma_1({\bf k})x_{\bf k}^2x_2^2\Big]
- \frac 1{2}\Big(\frac {2\eta}{1+\zeta} \Big) l_{\bf k}m_{\bf k}
{\tilde C}_{\bf k},
\label{C1C2}
\eea
with
\bea
{\tilde C}_{\bf k} &=& \Big(\frac {2}{N}\Big)^2 \sum_{12} l_1^2l_2^2\Big\{
\Big[2\gamma_2({\bf k})+\gamma_2(1)+\gamma_2(2)-4\gamma_2({\bf k}+1-2)\Big]x_1^2 \non \\
&+& \Big[\gamma_2(2)-\gamma_2(1+2-{\bf k})\Big](1+x_1^2x_2^2) \Big\}.
\label{Ceqn}
\eea

After defining the renormalized Green's function as in Ref.~\onlinecite{majumdar10b}
the first and second order self-energies are written as
\bea
\Sigma_{\alpha \alpha}^{(1)}({\bf k},\omega) &=& \Sigma_{\beta \beta}^{(1)}({\bf k},\omega)=A_{\bf k}, \\
\Sigma_{\alpha \beta}^{(1)}({\bf k},\omega) &=& \Sigma_{\beta \alpha}^{(1)}({\bf k},\omega)=B_{\bf k}, \\
\Sigma_{\alpha \alpha}^{(2)}({\bf k},\omega) &=& \Sigma_{\beta \beta}^{(2)}(-{\bf k},-\omega)
=C_{1{\bf k}} 
+ \Big(\frac {2}{N} \Big)^2\sum_{{\bf pq}}2l_{\bf k}^2l_{\bf p}^2l_{\bf q}^2l_{\bf k+p-q}^2 \non \\
&\times& \Big[\frac {|V^{(4)}_{\bf k,p,q,[k+p-q]}|^2}{\omega -E_{\bf p}
-E_{\bf q}-E_{\bf k+p-q}+i\delta} - \frac {|V^{(6)}_{\bf k,p,q,[k+p-q]}|^2}{\omega +E_{\bf p}
+E_{\bf q}+E_{\bf k+p-q}-i\delta}  \Big],\label{sigma1}\\
\Sigma_{\alpha \beta}^{(2)}({\bf k},\omega) &=& \Sigma_{\beta \alpha}^{(2)}(-{\bf k},-\omega)
=C_{2{\bf k}} 
+ \Big(\frac {2}{N} \Big)^2\sum_{{\bf pq}}2l_{\bf k}^2l_{\bf p}^2l_{\bf q}^2l_{\bf k+p-q}^2 {\rm sgn}(\gamma_{\bf G})\non \\
&\times& V^{(4)}_{\bf k,p,q,[k+p-q]}V^{(6)}_{\bf k,p,q,[k+p-q]}
\frac{2(E_{\bf p}+E_{\bf q}+E_{\bf k+p-q})}{\omega^2-(E_{\bf p}+E_{\bf q}+E_{\bf k+p-q})^2},
\label{sigma2}
\eea
where $[{\bf k+p-q}]$ is mapped to $({\bf k+p-q})$ in the first BZ by the reciprocal 
vector ${\bf G}$. 

The magnetization $M$ defined as the average of the spin operator
$S_z$ on a given sublattice (say A) is expressed as
\be
M = S-\langle a^\dag_i a_i \rangle = S-\Delta S + \frac {M_1}{(2S)}+\frac {M_2}{(2S)^2},
\label{Mag-AF}
\ee
where
\bea
\Delta S &=& \frac 1{N} \sum_{\bf k} \Big(\frac 1{\epsilon_{\bf k}}-1 \Big), 
\label{MagAF-LSWT} \\
M_1 &=& \frac 2{N} \sum_{\bf k}\frac {l_{\bf k}m_{\bf k}B_{\bf k}}{E_{\bf k}},  
\label{M1-AF} \\
M_2 &=& \frac 2{N} \sum_{\bf k} \Big\{ -(l_{\bf k}^2+m_{\bf k}^2)\frac {B_{\bf k}^2}{4E_{\bf k}^2}
+ \frac {l_{\bf k}m_{\bf k}}{E_{\bf k}}\Sigma^{(2)}_{\alpha \beta}({\bf k},-E_{\bf k}) \non \\
&-& \Big(\frac 2{N} \Big)^2 \sum_{\bf pq} 2l_{\bf k}^2l_{\bf p}^2l_{\bf q}^2l_{\bf k+p-q}^2 
\Big[\frac {(l_{\bf k}^2+m_{\bf k}^2)|V^{(6)}_{\bf k,p,q,[k+p-q]}|^2}
{(E_{\bf k}+E_{\bf p}+E_{\bf q}+E_{\bf k+p-q})^2} \non \\
&+& \frac {2l_{\bf k}m_{\bf k}{\rm sgn}(\gamma_{\bf G})V^{(4)}_{\bf k,p,q,[k+p-q]}
V^{(6)}_{\bf k,p,q,[k+p-q]}}{E_{\bf k}^2-(E_{\bf p}+E_{\bf q}+E_{\bf k+p-q})^2} \Big]
\label{M2-AF}
\Big\}.
\eea
The zeroth-order term $\Delta S$ corresponds to the reduction of magnetization within 
LSWT, $M_1$ term corresponds to the first-order $1/S$ correction, and $M_2$ is the second-order
correction.

\subsection{\label{sec:CAFphase}($\pi,0,\pi$) CAF Phase}
The Hamiltonian describing the CAF phase is
\bea
H &=& J_1 \sum_{i,\ell}{\bf S}_{i,\ell}^{\rm A} \cdot {\bf S}_{i+\delta_x,\ell}^{\rm B}
+ \half J_1^\prime\sum_{i,\ell}\Big[ {\bf S}_{i,\ell}^{\rm A} \cdot {\bf S}_{i+ \delta_y,\ell}^{\rm A}
+ {\bf S}_{i,\ell}^{\rm B} \cdot {\bf S}_{i+ \delta_y,\ell}^{\rm B}\Big]
+ J_2 \sum_{i,\ell}{\bf S}_{i,\ell}^{\rm A} \cdot {\bf S}_{i+\delta_x +\delta_y,\ell}^{\rm B} \non \\
&+& J_\perp \sum_{i,\ell}{\bf S}_{i,\ell}^{\rm A} \cdot {\bf S}_{i,\ell+1}^{\rm B}.
\label{ham-CAF}
\eea
The procedure is same as the AF phase. For this phase the structure factors $\gamma_{1{\bf k}}^\prime,\; 
\gamma_{2{\bf k}}^\prime$ along with other quantities required for the calculations are
defined as
\bea
\gamma^\prime_{1{\bf k}}&=&\big[\cos (k_x)(1+2\eta \cos (k_y))+\delta \cos (k_z)\big]/(1+2\eta), \non\\
\gamma^\prime_{2{\bf k}} &=& \cos(k_y), \non\\
\gamma^\prime_{\bf k} &=& \gamma^\prime_{1{\bf k}}/\kappa^\prime_{\bf k},\\
\kappa_{\bf k}^\prime &=& 1- \frac {\zeta}{1+2\eta} (1-\gamma^\prime_{2{\bf k}})+\frac{\delta}{1+2\eta},\non\\
\epsilon_{\bf k}^\prime &=& [1-\gamma_{\bf k}^{\prime 2}]^{1/2}. \non  
\eea
The coefficients that appear in the Hamiltonian $H_1$ are 
\bea
A_{\bf k}^\prime &=& A_1^\prime \frac 1{\kappa^\prime_{\bf k}\epsilon^\prime_{\bf k}}
\Big[\kappa^\prime_{\bf k}
-\gamma_{1{\bf k}}^{\prime 2}\Big] + A_2^\prime \frac 1{\epsilon^\prime_{\bf k}}
\Big[1-\gamma^\prime_{2 {\bf k}}\Big], \\
B_{\bf k}^\prime &=& B_1^\prime \frac {1}{\kappa^\prime_{\bf k}\epsilon^\prime_{\bf k}}
\gamma^\prime_{1 {\bf k}}\Big[1-\gamma^\prime_{2{\bf k}}\Big],
\label{definitions-CAF} 
\eea
with
\bea
A_1^\prime &=& \frac 2{N} \sum_{\bf p} \frac 1{\epsilon^\prime_{\bf p}}
\Big[\frac {\gamma_{1{\bf p}}^{\prime 2}}{\kappa^\prime_{\bf p}}+\epsilon^\prime_{\bf p}-1\Big], \\
A_2^\prime &=& \Big(\frac {\zeta}{1+2\eta} \Big)\frac 2{N} \sum_{\bf p} 
\frac 1{\epsilon^\prime_{\bf p}}\Big[1-\epsilon^\prime_{\bf p}-\gamma^\prime_{2{\bf p}}\Big], \\
B_1^\prime &=& \Big(\frac {\zeta}{1+2\eta} \Big)\frac 2{N} \sum_{\bf p} 
\frac 1{\epsilon^\prime_{\bf p}}\Big[\gamma^\prime_{2{\bf p}}-
\frac {\gamma_{1{\bf p}}^{\prime 2}}{\kappa^\prime_{\bf p}}\Big].
\eea
$H_0,\;H_1$, and $H_2$ can be expressed in the same forms as in Eqs.~\ref{H0term},~\ref{H1term},
and ~\ref{H2term} with the new coefficients $A_{\bf k}^\prime, B_{\bf k}^\prime,C_{1{\bf k}}^\prime, 
C_{2{\bf k}}^\prime$ and with the replacement $\zeta \leftrightarrow 2\eta$. 
The expressions for the two vertex factors $V^{\prime (4)}, V^{\prime (6)}$ and the 
coefficients $C_{1{\bf k}}^\prime, C_{2{\bf k}}^\prime $ are similar to the AF phase (details can be 
found in Ref.~\onlinecite{majumdar10b}).

\section{\label{sec:results}Magnetization and the Phase diagram}

\begin{figure}[httb]
\centering
\includegraphics[width=3.8in]{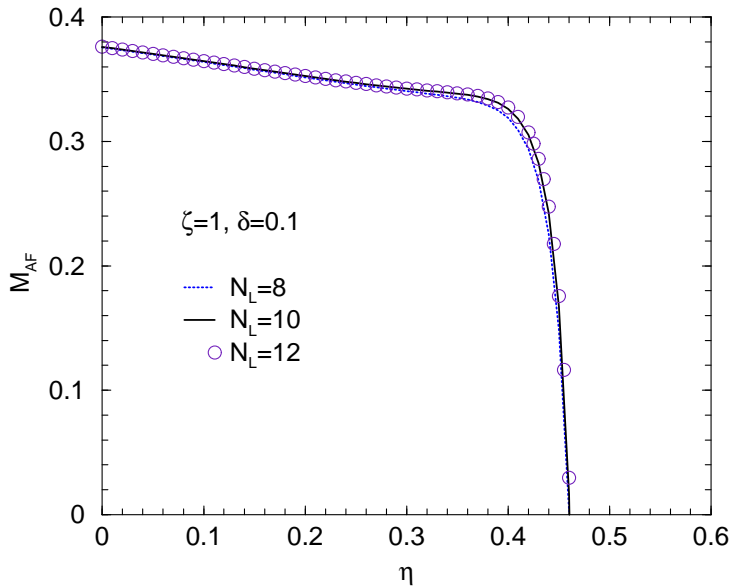}
\caption{\label{fig:Mconv} (Color online) Convergence test of magnetization calculation for the AF phase ($\zeta=1, \delta=0.1$) with $N_L=8$ (dashed line), 10 (solid line), and 12 (open circles) lattice sites . The convergence is excellent with $N_L=12$ sites.}
\end{figure}

We obtain the sublattice magnetization $M$ for the
two ordered phases with different values of $\zeta$,
$\eta$, and $\delta$ from Eq.~\ref{Mag-AF} by numerically evaluating Eqs.~\ref{MagAF-LSWT}--\ref{M2-AF}
(using similar expressions for the CAF phase). 
Especially to obtain the second order correction term $M_2$ we sum
up the values of $N_L^3/8$ points of ${\bf k}$ in the 1/8-th part of the 
first BZ and $N_L^3$ points of ${\bf p}$ and $N_L^3$ points of ${\bf q}$ in the full BZ.
To check the convergence of our results we do the calculations for the AF-phase with $N_L=8, 10$, and 12 sites for $\zeta=1$ and $\delta=0.1$. The convergence is very good as shown in Fig.~\ref{fig:Mconv}. $M_{\rm AF}$ becomes zero at the critical point $\eta_{1c} \approx 0.460$. Hereafter, we use $N_L=12$ lattice sites for all of our numerical computations. Evaluation of the magnetization requires summing contributions from over 645 million points in the first BZ for each $\zeta, \eta$, and $\delta$.

\begin{figure}[httb]
\centering
\includegraphics[width=3.4in]{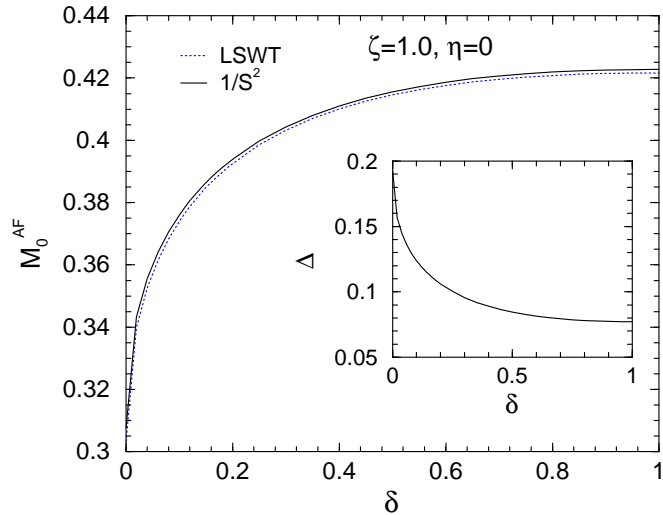}
\caption{\label{fig:M0AF} (Color online) Magnetization $M_{0}^{\rm AF}$ for the AF phase plotted for the unfrustrated case ($\eta=0$) with interlayer coupling $\delta$ for $\zeta=1.0$. Dotted line is the result from LSWT and the solid line is with $1/S^2$ corrections. $M_{0}^{\rm AF}$ (with second-order corrections) increases from 0.308 for $\delta=0$ to 0.423 for $\delta=1.0$. Inset shows the spin-deviation $\Delta = 0.5-M_{0}^{\rm AF}$.  With increase in $\delta$ the fluctuations decrease, thus $M_{0}^{\rm AF}$ increases.  This result is expected as with increase in the interlayer coupling the system makes a transition from two to three dimensions.}
\end{figure}

Figure \ref{fig:M0AF} shows the sublattice magnetization $M_{0}^{\rm AF}$ for the AF phase with $\zeta=1$ and $\eta=0$ for  different values of interlayer coupling $\delta$. In the inset we plot the spin-deviation $\Delta = 0.5-M_{0}^{\rm AF}$ with $\delta$. $\Delta$ is a measure of quantum fluctuations from the classical value of 0.5. We find that with 
increase in $\delta$ the fluctuations decrease, thus $M_{0}^{\rm AF}$ (with second-order corrections) increases from 0.308 for $\delta=0$ to 0.423 for $\delta=1.0$. This result is expected as with the increase in interlayer coupling the system undergoes a dimensional transition from 2D to 3D. The dotted line is the result from LSWT. Result for $M_{0}^{\rm AF}$ obtained from LSWT captures the essential
physics both qualitatively and quantitatively as second-order corrections are small for the unfrustrated ($\eta=0$) case.

\begin{figure}[httb]
\centering
\includegraphics[width=3.0in]{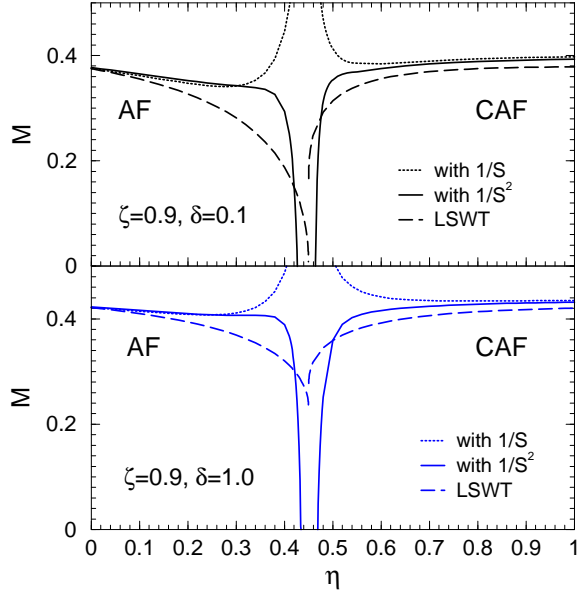}
\caption{\label{fig:MagAF-CAF1} (Color online) Sublattice magnetization $M$ for the AF and CAF ordered phases for spatial anisotropy $\zeta=0.9$ with two different values of interlayer couplings $\delta=0.1$ and 1.0. For each $\delta$ results from LSWT (long-dashed lines), with first-order $1/S$ corrections (dotted lines), and with second-order $1/S^2$ corrections (solid lines) are plotted. With increase in $\eta$ dotted lines diverge. Second-order corrections become significant with increase in $\eta$ and stabilize the 
anomalous divergence of the magnetization. LSWT calculations for $\delta=0.1$ show that the magnetization becomes zero for the AF phase at $\eta_{1c}\approx 0.45$ whereas it does not go to zero from the CAF phase. LSWT is not applicable at the classical transition point $\eta_c^{\rm class}=0.45$. Extrapolation of the CAF phase LSWT results show a direct transition from the AF to the CAF phase.  For $\delta=1.0$, $M$ never becomes zero for both the phases (within LSWT) indicating that there is a direct first-order phase transition from the AF to the CAF phase. However, with second-order corrections $M$ becomes zero for both the phases at the critical points 
$\eta_{1c} \approx 0.427, \eta_{2c} \approx 0.464$ for $\delta=0.1$ and $\eta_{1c} \approx 0.434, \eta_{2c} \approx 0.469$ for $\delta=1.0$.}
\end{figure}

\begin{figure}[httb]
\centering
\includegraphics[width=3.0in]{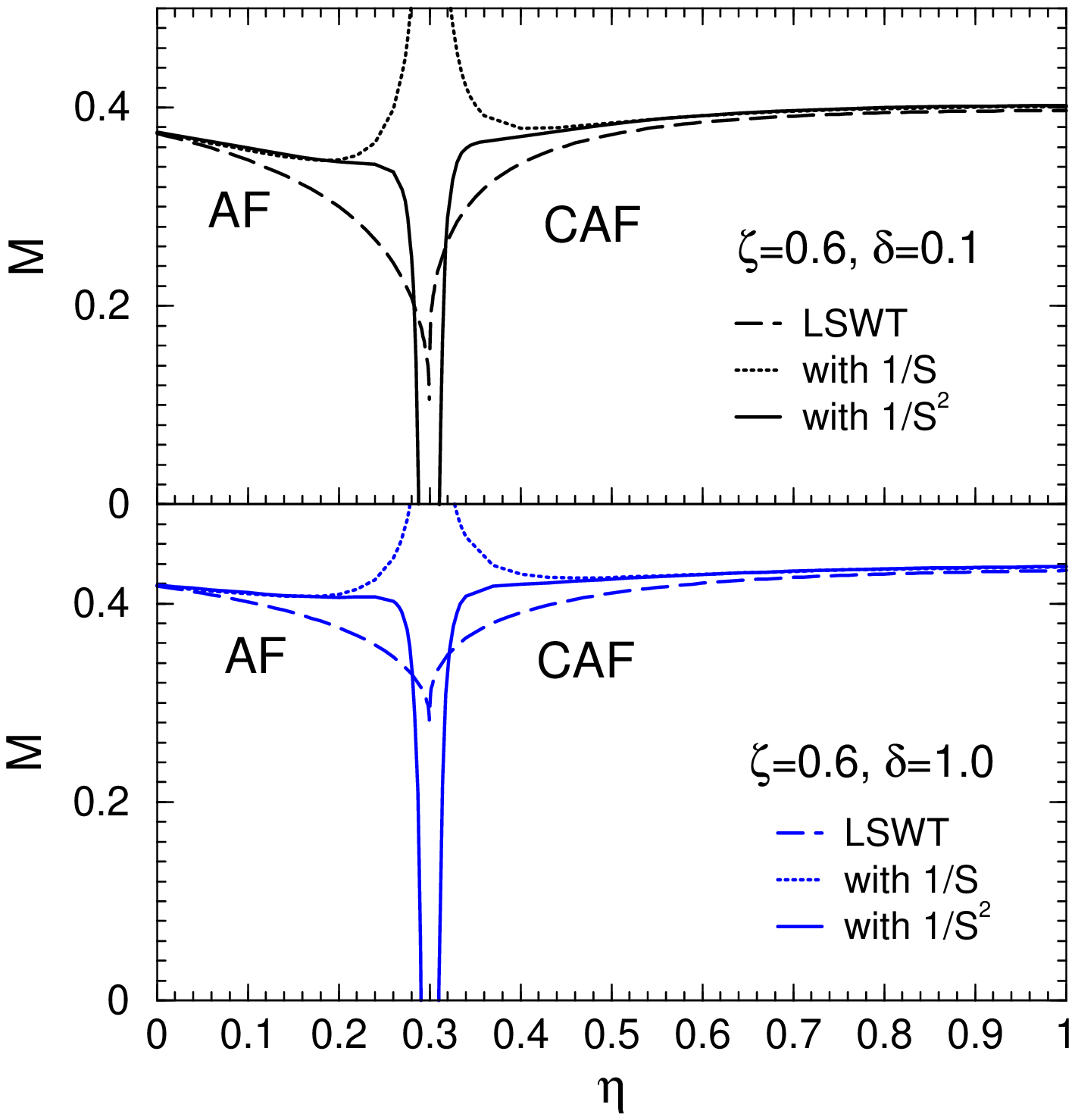}
\caption{\label{fig:MagAF-CAF2} (Color online) Magnetization $M$ for the AF and CAF ordered phases shown
for spatial anisotropy $\zeta=0.6$ with interlayer couplings $\delta=0.1$ and 1.0. Similar to Fig.~\ref{fig:MagAF-CAF1} $M$ with $1/S^2$ corrections decrease steadily and then sharply drops to zero for both the phases. LSWT calculations for both $\delta=$0.1 and 1.0  show that there is a direct first-order phase transition from the AF to the CAF phase. However, with second-order corrections we find that $M$ vanishes for both the phases at the critical points $\eta_{1c} \approx 0.288, \eta_{2c} \approx 0.311$ for $\delta=0.1$ and $\eta_{1c} \approx 0.290, \eta_{2c} \approx 0.310$ for $\delta=1.0$.}
\end{figure}

Figures~\ref{fig:MagAF-CAF1} and \ref{fig:MagAF-CAF2} show the magnetization with
increase in the frustration parameter $\eta$ for two different values of the spatial
anisotropy parameter $\zeta=0.9$ and 0.6.  Similar to the 2D case (see Ref.~\onlinecite{majumdar10b} for details) our spin-wave expansion for the CAF phase becomes unreliable as $\zeta$ gets close to 1. Thus we have not chosen $\zeta=1$ for our plots. For each $\zeta$ and $\delta$ three different curves are plotted: the long-dashed lines represent LSWT results, the dotted lines include the first-order $1/S$ corrections, and the solid lines represent corrections up to second-order to the LSWT results. As $\eta$ approaches the classical transition point $\eta^{\rm class}_c=0.45$ from both sides of the two ordered phases the dotted curves diverge. However, $1/S^2$ corrections ($M_2$) significantly increase to stabilize the divergence. We find that the magnetizations with $1/S^2$ corrections decrease steadily and then sharply drops to zero for both the phases as $\eta \rightarrow \eta^{\rm class}_c$. As an example, with $\zeta=0.9,\; \delta=0.1$ magnetization for the AF phase begins from $\approx 0.377$ at $\eta=0$, then  decreases till $\eta \approx 0.41$, and sharply becomes zero at the critical point $\eta_{1c} \approx 0.427$. $M_2$ corrections start from a small positive number at $\eta=0$ and then switches sign at $\eta \approx 0.32$. $M$ for the CAF phase shows the same feature, where we find $M \approx 0.393$ at $\eta=1$ and $\eta_{2c} \approx 0.464$. However, in this case $M_2$ corrections are always negative. These results are qualitatively similar to the results for the 2D spatially anisotropic, frustrated HAFM on a square lattice ($\delta=0$).\cite{majumdar10b}

On the other hand, LSWT calculations for $\zeta=0.9$ and $\delta=0.1$ show that the magnetization becomes zero for the AF phase at $\eta_{1c}\approx 0.45$ whereas it does not go to zero from the CAF phase. LSWT is not applicable at the classical transition point $\eta=0.45$. Extrapolation of the CAF phase LSWT results show a direct transition from the AF to the CAF phase.

For $\zeta=0.9$ and $\delta=1.0$, $M$ never becomes zero for both the AF and CAF ordered phases (within LSWT) indicating that there is a direct first-order phase transition from the AF to the CAF phase. But with second-order corrections we find that $M$ becomes zero for both the phases at the critical points $\eta_{1c} \approx 0.434$ and $ \eta_{2c} \approx 0.469$. With increase in $\delta$ the width of the disordered PM phase diminishes but it survives even for large $\delta$. 

In Fig.~\ref{fig:MagAF-CAF2} we show the results for $\zeta=0.6$. LSWT calculations show a first-order direct phase transition from AF to the CAF phase for both $\delta=0.1$ and 1.0. Similar to the 
$\zeta=0.9$ case we find that $M$ vanishes (with second-order corrections) for both the phases at the critical points $\eta_{1c} \approx 0.288, \eta_{2c} \approx 0.311$ for $\delta=0.1$ and $\eta_{1c} \approx 0.290, \eta_{2c} \approx 0.310$ for  $\delta=1.0$. The two phases are separated by a narrow disordered paramagnetic region. We repeat the calculations for $\zeta=0.4$ with different values of $\delta$ and obtain similar features (the results are not shown).

\begin{figure}[httb]
\centering
\includegraphics[width=3.0in]{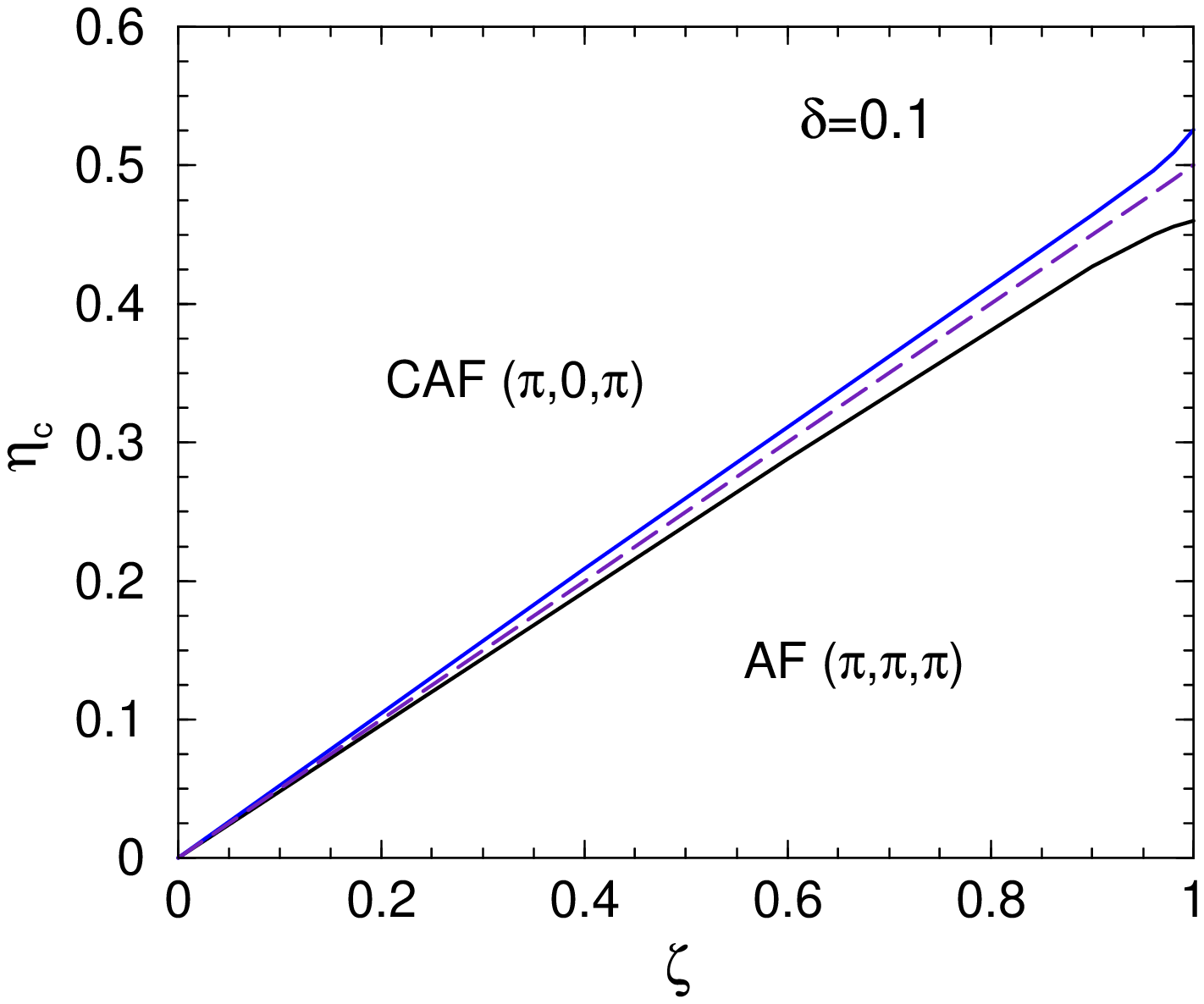}
\caption{\label{fig:phase-zeta} (Color online) Phase diagram for the $J_1- J_1^\prime- J_2- J_\perp$ model with $\delta=0.1$. Solid lines represent the critical phase transition points $\eta_{1c}$ and $\eta_{2c}$ for the AF and CAF phases. Dashed line shows the classical phase transition line $\eta_c^{\rm class}=\zeta/2$. For the AF phase  $\eta_{1c} \approx 0.460$ at $\zeta=1$. Spin-wave expansion for the CAF phase becomes unreliable for $\zeta$ near 1. We thus extrapolate our data to obtain $\eta_{2c}\approx 0.53$ for $\zeta=1$. The width of the disordered region $(\eta_{2c}-\eta_{1c})$ increases with the anisotropy parameter $\zeta$.}
\end{figure}

Another feature we observe from our data is that with decrease in spatial anisotropy $\zeta$ the width of the disordered region $(\eta_{2c}-\eta_{1c})$ diminishes but it never disappears. This is shown in 
Fig.~\ref{fig:phase-zeta} for $\delta=0.1$. The solid lines represent the critical points $\eta_{1c}$ and 
$\eta_{2c}$ for the AF and CAF phases. The dashed line is the classical first-order phase
transition line $\eta_c^{\rm class}=\zeta/2$ (independent of $\delta$) between the two phases. For the AF phase $\eta_{1c} \approx 0.460$ for $\zeta=1$. For the CAF phase we obtain the value of the critical phase transition point $\eta_{2c}\approx 0.53$ at $\zeta=1$ by extrapolation as our spin-wave expansion is unreliable for $\zeta$ near 1. These results can be compared to the results for the spin-$\half$ spatially anisotropic, frustrated HAFM on a square lattice ($\delta=0$) where we have found $\eta_{1c}\approx 0.41$ and $\eta_{2c}\approx 0.58$ for $\zeta=1$.\cite{majumdar10b}. All other features are similar to the phase diagram of the anisotropic, frustrated 2D model.

\begin{figure}[httb]
\centering
\includegraphics[width=3.0in]{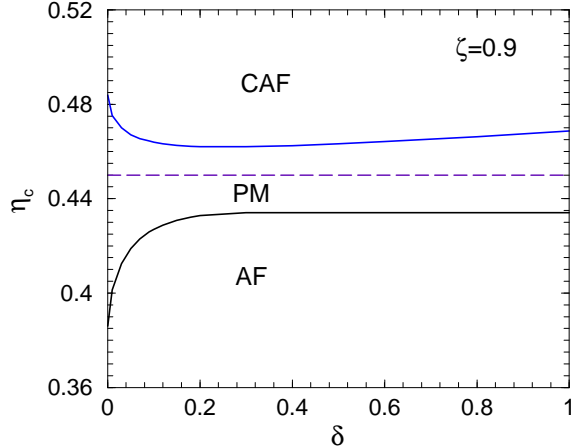}
\caption{\label{fig:phasediagram} (Color online) Phase diagram for the $J_1- J_1^\prime- J_2- J_\perp$ model with $\zeta=0.9$. The dashed line shows the classical phase transition line $\eta_c^{\rm class}=0.45$. Phase boundaries of the two ordered phases AF and CAF diminishes but never disappears with increase in 
$\delta$. These two ordered phases are always separated by the magnetically disordered phase. For the AF phase $\eta_{1c}$ increases from $\approx 0.386$ for $\delta=0$ to $\approx 0.434$ for $\delta = 1.0$. 
Whereas for the CAF phase $\eta_{2c}$ begins from $\approx 0.484 $ and then slowly decreases to $\approx 0.462$ for $\delta =0.3$. But from $\delta \approx 0.4$ the phase boundary for the CAF phase shows a slight upward rise. By excluding the $J_2$ interactions in the $xz$ and $yz$ planes our model becomes unrealistic  in the strong interlayer coupling limit where $J_\perp$ is comparable to $J_1$ and $J_1^\prime$ (especially in the CAF phase).}
\end{figure}

Phase diagram for the $J_1-J_1^\prime-J_2-J_\perp$ model for $\zeta=0.9$ is displayed in 
Fig.~\ref{fig:phasediagram}. Within our spin-wave expansion we do not find any critical value of 
$J_\perp$ above which there is no disordered region. This is in contrary to the findings in Refs.~\onlinecite{darradi06} and \onlinecite{nunes10}. Instead from our calculations we find 
that the transition between the two phases is always 
separated by the magnetically disordered phase. For the AF phase $\eta_{1c}$ increases from $\approx 0.386$ for $\delta=0$ to $\approx 0.434$ for $\delta = 1.0$. On the other hand for the CAF phase $\eta_{2c}$ begins from $\approx 0.484 $ and then slowly decreases to $\approx 0.462$ for $\delta =0.3$.
Then from $\delta \approx 0.4$ the phase boundary for the CAF phase shows a slight upward rise. In this strong interlayer coupling limit where $J_\perp $ becomes comparable to $J_1$ and 
$J_1^\prime$ for the CAF phase ($J_2 >J_1$) our model becomes unrealistic as we have excluded the NNN interactions in the $xz$ and $yz$ planes. Our present model differs from the spin-$\half$ Heisenberg AF on a fully frustrated simple cubic lattice where we have the additional $J_2$ interactions in the $xz$ and $yz$ planes. In that case it has been shown that there is a direct first-order phase transition from the AF to the CAF phase.~\cite{majumdar10a}

\section{\label{sec:conclusions}Conclusions}
In this work using second-order spin-wave expansion we have studied the effects of interlayer
coupling and directional anisotropy on the magnetic phase diagram of a frustrated spin-$\half$ Heisenberg antiferromagnet on a stacked square lattice. Linear spin-wave theory calculations for this model 
show that for small interlayer coupling there are two magnetic ordered phases,  AF and CAF which are separated by a disordered paramagnetic state. However when the interlayer coupling exceeds a critical value the disordered paramagnetic phase disappears and then there is a direct first-order phase transition from the AF to the CAF phase. Recent numerical calculations using coupled cluster and rotation-invariant Green's function methods support this picture.\cite{darradi06,nunes10} With our second-order spin-wave expansion we have found that with increase in next nearest neighbor frustration, $1/S^2$ corrections play a significant role in stabilizing the magnetization as the classical phase transition point is approached. As expected from linear spin-wave theory we have found that there are two ordered magnetic phases (Ne\'{e}l and stripe) which are separated by a paramagnetic disordered phase. But the values of the critical phase transition points for the two phases differ from the LSWT predictions. Our calculations show that the width of the disordered region diminishes with decrease in the directional anisotropy.  These features are similar to the magnetic phase diagram of a two-dimensional frustrated Heisenberg spin-$\half$ antiferromagnet.\cite{majumdar10b} However, with increase in the interlayer coupling we have found that the parameter region of this disordered phase does not disappear. Our obtained phase diagram is significantly different from the phase diagram obtained using linear spin-wave theory which predicts a direct first order phase transition from the AF to the CAF phase beyond a critical value of interlayer coupling. In summary with our present approach based on second-order spin wave expansion we do not find existence of any critical interlayer coupling (or spatial anisotropy) beyond (or below) which there is a direct transition from one phase to the other ordered phase. 

\section{Acknowledgments}
The author thanks H. Johannesson for useful discussions and suggestions. This project acknowledges the use of the Cornell Center for Advanced Computing's ``MATLAB on the TeraGrid'' experimental computing resource funded by NSF grant 0844032 in partnership with Purdue University, Dell, The MathWorks, and Microsoft.

\bibliography{Aniso3D}

\end{document}